# An Explorative Study on Document Type Assignment of Review Articles in Web of Science, Scopus and Journals' Website

Manman Zhu[1,2,3,‡], Xinyue Lu[1,2,‡], Fuyou Chen[1], Liying Yang[1,2], Zhesi Shen[1,*—]

1. National Science Library, Chinese Academy of Sciences, Beijing 100190, P. R. China;

2. Department of Information Resource Management, School of Economics and Management, University of Chinese Academy of Sciences, Beijing 100190, P. R. China

3. Institute of Science and Technology Information, Jiangsu University, Zhenjiang, 212013, P. R. China;

**Abstract:**

Accurately assigning the document type of review articles in citation index databases like Web of Science(WoS) and Scopus is important. This study aims to investigate the document type assignation of review articles in web of Science, Scopus and Journals' website in a large scale. 27,616 papers from 160 journals from 10 review journal series indexed in SCI are analyzed. The document types of these papers labeled on journals' website, and assigned by WoS and Scopus are retrieved and compared to determine the assigning accuracy and identify the possible reasons of wrongly assigning. For the document type labeled on the website, we further differentiate them into explicit review and implicit review based on whether the website directly indicating it is review or not. We find that WoS and Scopus performed similarly, with an average precision of about 99% and recall of about 80%. However, there were some differences between WoS and Scopus across different journal series and within the same journal series. The assigning accuracy of WoS and Scopus for implicit reviews dropped significantly. This study provides a reference for the accuracy of document type assigning of review articles in WoS and Scopus, and the identified pattern for assigning implicit reviews may be helpful to better labeling on website, WoS and Scopus.

**Keywords:** Document Type; Web of Science; Scopus; Review Article

## 1. Introduction

### 1.1 Reviews play an importance role in science and science communication

As Garfield (1996) described,

*"Reviews play an essential role in scientific communication and understanding. In terms of the inherent characteristics of the review, they can provide a synthesis of the proliferating fragmented knowledge appearing in the plethora of foreign and domestic journals in a*

*corresponding author：Zhesi Shen (shenzhs@mail.las.ac.cn)

‡ These authors contribute equally

*specialty or subspecialty. As such, they can elucidate trends in research and point to unanswered questions that provide opportunities for future study. Reviews also give science policymakers as well as researchers a clearer insight into the potential importance of emerging knowledge."*

In addition, review provides excellent and stimulating reading for the general reader and researcher dedicated to cross-disciplinary study, because they advance our perceptions of the relationships between different research efforts. The value of a review does not exist solely in the author's synthesis of previously published papers; the bibliography in a review usually is a high quality list of core articles about the subject. In all, writing a review will certainly do as much for the advancement of science as those do the original research.

## 1.2 The necessity of accurately assigning document type of reviews in databases

Despite the importance of reviews in science and science communication, the effect of reviews on scientometrics analysis is also significant. Reviews tend to be more frequently cited(Aksnes, 2003; Moed, 2010; Teixeira et al., 2013). Correlated with this overcitation, there is an overrepresentation of reviews in the highly cited papers, and this overrepresentation becomes greater when the most highly cited papers are considered(Miranda et al., 2018). Moreover, 20% reviews can increase the average citations of an individual researcher with 40%–80%. Consequently, researchers boost their citation by publishing reviews, and journals increase their Impact Factor by publishing reviews(Ketcham et al., 2007; Teixeira et al., 2013; Lei et al., 2020).

Review will also affect the citation of the articles it reviewed. An alarming trend within the biological/biomedical science had been noted that the authors prefer to cite review articles rather than the original article when writing literature review(Ketcham et al., 2007; Teixeira et al., 2013). It is more efficient to cite reviews than all the individual studies, but the scientific credit to the time-consuming original studies will be absorbed by the reviews and the review authors(Ketcham et al., 2007; Teixeira et al., 2013; Lachance et al., 2014). Ho et al. (2017)pointed that review papers will affect the main path analysis and clustering analysis. When conducting bibliometrics research and evaluating scientific research achievements, we should decide which document type to be included and whether treat articles of different document types separately(Lei et al., 2020). To facilitate the above process, highly accurate assignment of review articles in databases is required. Wrongly assigned document type has great impact on the citation-based evaluation(Donner, 2017; Zhu et al., 2022).

## 1.3 Definition of review in databases and related studies of the document type assignment of reviews in databases

WoS describes the Review$^{-}$ as

*"Detailed, critical surveys of published research. A review article may summarize previously published studies and draw some conclusions on the subject. Includes Reviews, Review of Literature, Mini-reviews, and Systematic reviews. If an*

*article is listed under the review section in a journal and/or Review of Literature appears in the title it will be assigned a review.*

*If an article is not assigned a review by the journal but Review, Systematic Review or Mini-review appears in the title, it must also appear someplace else in the article (abstract/summary or introduction) in order to be assigned the document type review.*

*NOTE: If the article(s) meet the above criteria - they must have References in order to be tagged as a Review item.*

*Review articles that were presented at a Symposium or Conference will be processed as Proceedings Papers."*

This description is accessed recently. Several years before, the description of "Review" in WoS is a renewed study or survey of previously published literature providing new analysis or summarization of the research topic(WoS, 2023). And several criteria are used to determine whether a paper is a review. such as the following:

*"In the JCR system any article containing more than 100 references is coded as a review. Articles in 'Review' sections of research or clinical journals are also coded as reviews, as are articles whose titles contain the word 'Review' or 'overview'"(Garfield, 1994).*

The "more than 100 references" criteria had been removed in 2010. The change of criteria may lead to some change to the statistics. For example, in a 1987 paper, Eugene Garfield pointed out that there were 625432 articles indexed in the 1986 SCI and of which approximately 32000 are reviews(Garfield, 1987). But when we search the SCI in June 2021, total articles indexed in 1986 are 709136 and 13197 are review.

Scopus describes Review as

"A significant review of original research also includes conference papers. Reviews typically have an extensive bibliography. Educational items that review specific issues within the literature are also considered to be reviews. As non-original articles, reviews lack the most typical sections of original articles such as materials & methods and results"(McCullough, 2023).

In Scopus, there is another document type related to review called "Short survey", which is described as

*"Short or mini-review of original research. Short surveys are similar to reviews, but usually are shorter (not more than a few pages) and with a less extensive bibliography."*

We can see that the description of reviews in Scopus and WoS is mainly related to the words used in titles and abstracts, the length of the reference list and article structure. Colebunders et al. (2013) compared the number of records related to reviews retrived in WoS via different strategies, i.e., (1) based on the WoS document type, (2) having either the word review or the word overview in the title, and (3) a topic search (TS=) for the words review or reviews. It is found that the absolute and relative numbers of reviews differ depending on which of the three definitions are used. Harzing (2013) reported a comprehensive analysis of document categories for 27 journals in nine Social Science and Science disciplines and showed that WoS may misclassified social science journal articles containing original research into the 'review' or 'proceedings paper' category. The possible reason is length of references of social science articles is larger than 100.

Several studies compared the document type assignment accuracy of citation index database with other sources(e.g., manually coded, publisher's website). Hayashi et al. (2013) compared the records' document type of 18 research journals of Nature Publishing Group in WoS, Scopus, and the website and found that all "Review" items in the website were labeled as "Review" in both WoS and Scopus, and some papers of other types are also labeled as "Review" in WoS and Scopus. As the authors didn't further report the details of these reviews labeled in WoS and Scopus, we cannot infer the real accuracy. Donner (2017) reported a study on the document type assignment accuracy of 791 randomly selected papers in WoS and Scopus. When only focusing on these selected papers, the accuracy of WoS(83%) is higher than Scopus(76%). The study also statistically inferred the WoS overall proportion of correctly assigned DT is 0.94, but for the reviews, the precision is 0.87 and recall is 0.57. Yeung (2019) examined the DT assignment accuracy of 400 top cited publication defined by Scopus as 'article' in the field of food and nutritional sciences. Among these 400 publications, 117 were manually coded as review. Further, for these 117 reviews, 111 is indexed in WoS and 55/111 were wrongly labeled. Another interesting observation is that, the publisher website labeled 52/117 reviews as article.

Reviews have always been a very important research object in the field of Scientometrics and Informetrics, and the research directions about reviews had been discussed at a workshop in the Conference of the International Society of Informetrics in September 2019, in which participants identified six realms of study. One of the themes is "the study of methodological caveats resulting from the usage of scholarly databases", such as lack of accuracy of document assignation in scholarly data bases(Blümel et al., 2020). In this work, we'd like to analyze the accuracy of document type assignment of review articles in WoS and Scopus in a large scale and identify the possible reasons of wrongly assigning.

## 2. Data and Methods

### 2.1 Data collection

In the publishing ecosystem, there are several journal series mainly (or only) publishing review articles, e.g., the Annual Reviews series, Nature Reviews series. These journals can be treated as appropriate data source for us to investigate the correctness of document type assignment of reviews in databases. For example, as shown in Fig.1, a paper published by *Nature Reviews Cancer*, has a document type annotation on its official website, and can be further compared with the corresponding document type provided in WoS and Scopus. In this study, we selected 160 SCI journals included in Journal Citation Report 2019 from five series of pure review journals (only publishing review articles) and four series of mixed review journals(mainly publishing review articles) as shown in Table1.

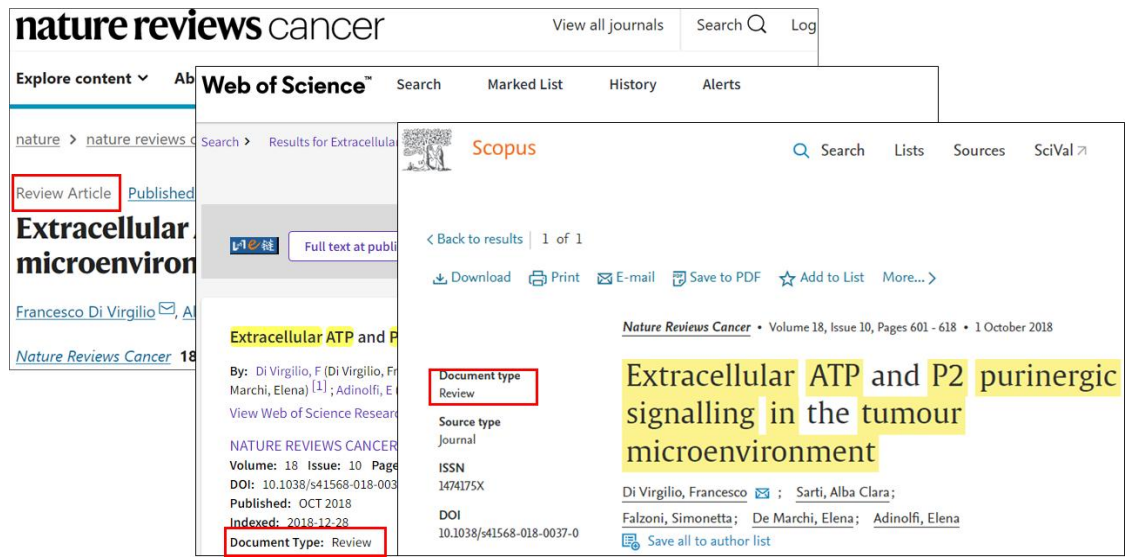

**Figure 1.** An example of the document type annotation of a review paper on its official Website, Web of Science and Scopus.

**Table 1.** List of review journal series investigated

| Series of review journal | Type of Review Journal | NO. of journals | NO. of papers |
|---|---|---|---|
| Annual Reviews | Pure | 39 | 1842 |
| Cell Trends In series | Pure | 15 | 3206 |
| Wolters Kluwer Current Opinion | Pure | 24 | 4333 |
| Reviews of Modern Physics | Pure | 1 | 86 |
| WIREs-Wiley Interdisciplinary Reviews | Pure | 9 | 755 |
| Elsevier Current Opinion | Mixed | 20 | 4983 |
| Nature Reviews | Mixed | 18 | 5975 |
| Taylor & Francis Expert Opinion | Mixed | 11 | 2519 |
| Taylor & Francis Expert Review | Mixed | 13 | 2737 |
| Taylor & Francis Critical Review | Mixed | 10 | 1180 |
| Total | - | 160 | 27616 |

As JCR 2019 covering papers published during 2017-2018, the website annotation information of papers published in the above review journals during 2017-2018 are collected from the journals' official websites as the basic dataset. We collect 27,616 papers, and for these collected papers, we retrieved their document type information from WoS and Scopus using Digital Object Identifier(DOI). Because of the problems such as record missing, errors or duplicate DOI in WoS and Scopus, we use the paper title, journal name and other supplementary information to manually match unmatched records.

## 2.2 Measurement of assignment accuracy

For pure review journals and mixed review journals, the document type of each paper is assigned based on the journal section headings or the document type annotation on the paper's official website, see the examples shown in Fig.2. As we mainly focus on the assignment of reviews and different journals and database have different names for the same document type, to facilitate the analysis, we grouped these document type names into "reviews", "articles" and "other papers"(e.g., editorial meterial, correction). We further divide the reviews/articles into explicit review/article (the section heading or the website annotation directly indicates its document type) and implicit review/article (the section heading or the website annotation doesn't directly indicate its document type). The details of the division are described for each journal series in the Results section. "Short survey" is a special document type in Scopus, and we keep its original name.

To measure the assignment accuracy of WoS and Scopus compared against official website, we construct an assignment matrix and calculate the corresponding precision, recall, and $F_1$-score metrics(Baeza-Yates et al., 1999; Davis et al., 2006) as follows,

$$\text{Precision} = N^{\text{web}}_{\text{db}} / N_{\text{db}}$$

$$\text{Recall} = N^{\text{web}}_{\text{db}} / N^{\text{web}}$$

$$F_1\text{-Score} = 2/ (1/ \text{Precision} + 1/ \text{Recall})$$

where $N^{\text{web}}_{\text{db}}$ is the number of papers marked as review both in website and database, $N_{\text{db}}$ is the number of papers marked as review in database, and $N^{\text{web}}$ is the number of papers marked as review on website.

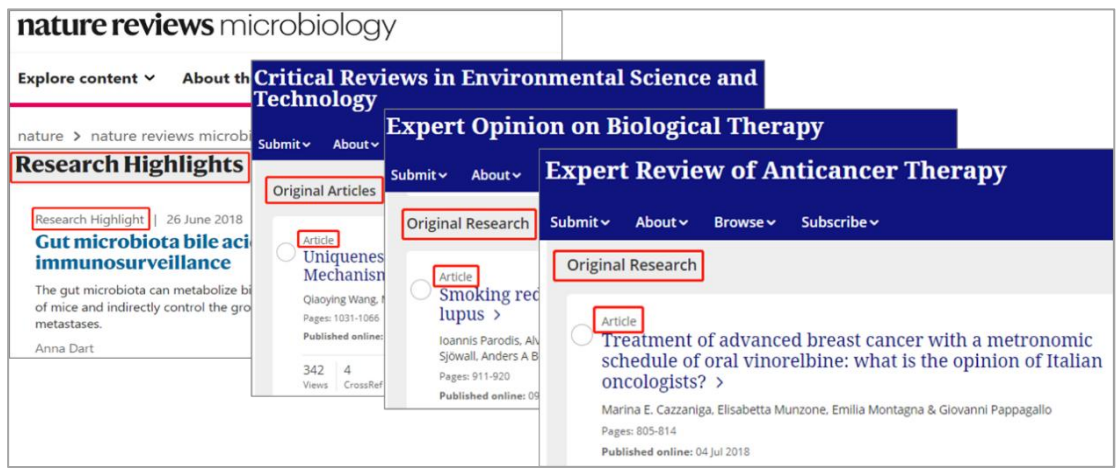

**Figure 2.** Examples of section headings and document type annotations for mixed review journals

# 3. Result

In this section we will present the comparisons of each journal series, and a graphical illustration of the document type correspondence of website and databases can be found in appendix.

## 3.1 Descriptive results of review mark for pure review journals

### 3.1.1 Annual Review journal series

As described on the website, Annual Reviews series are pure review journals, which

➢ *Capture current understanding of a topic, including what is well supported and what is controversial;*
➢ *Set the work in historical context;*
➢ *Highlight the major questions that remain to be addressed and the likely course of research in upcoming years;*
➢ *Outline the practical applications and general significance of research to society.*

Due to there are no section heading on website, papers that not titled as "Introduction", "Related articles" or other editorial material like names are assigned as "explicit review".

Table 2 shows the assignment result for Annual Reviews. 1501(83.62%) explicit reviews are labeled as "Review" in WoS, and 1285(71.59%) are labeled as "Review" in Scopus. Some papers entitled as "Introduction" and "Related articles" are not indexed by WoS and Scopus.

**Table 2** Assignment matrix for Annual Reviews series

| Type | Annual Reviews | Total | Web of Science | | | | Scopus | | | | |
|---|---|---|---|---|---|---|---|---|---|---|---|
| | | | review | article | others | not indexed | review | article | short survey | others | not indexed |
| Review | Explicit | 1795 | 1501 | 292 | 2 | 0 | 1285 | 505 | 0 | 5 | 0 |
| Other paper | | 47 | 2 | 0 | 22 | 23 | 4 | 2 | 1 | 31 | 9 |
| Total | | 1842 | 1503 | 292 | 24 | 23 | 1289 | 507 | 1 | 36 | 9 |

When we investigate the 292 misassigned reviews in WoS, we find that they are from seven journals, in which all the explicit reviews published by *Annual Review of Cancer Biology*, *Annual Review of Clinical Psychology*, *Annual Review of Virology* and *Annual Review of Analytical Chemistry* are labeled as "Article".

Scopus correctly labeled all the reviews published by 8 journals, e.g., *Annual Review of Analytical Chemistry*, *Annual Review of Biophysics*, etc. But for ten journals, more than half of reviews are labeled as "Article" in Scopus, with *Annual Review of Virology* having the largest proportion(78.72%).

### 3.1.2 Cell Trends In journal series

On the website of Cell Trends In series journals, the document type is annotated above the title of each paper as shown in Fig.3. Papers with annotation "Review" are marked as "explicit review", papers with annotation "mini review" are marked as 'mini review', and papers of the other 5 categories ("Correspondence", "Discussion", "Book Review", "Erratum", "Editorial") are marked as "other paper".

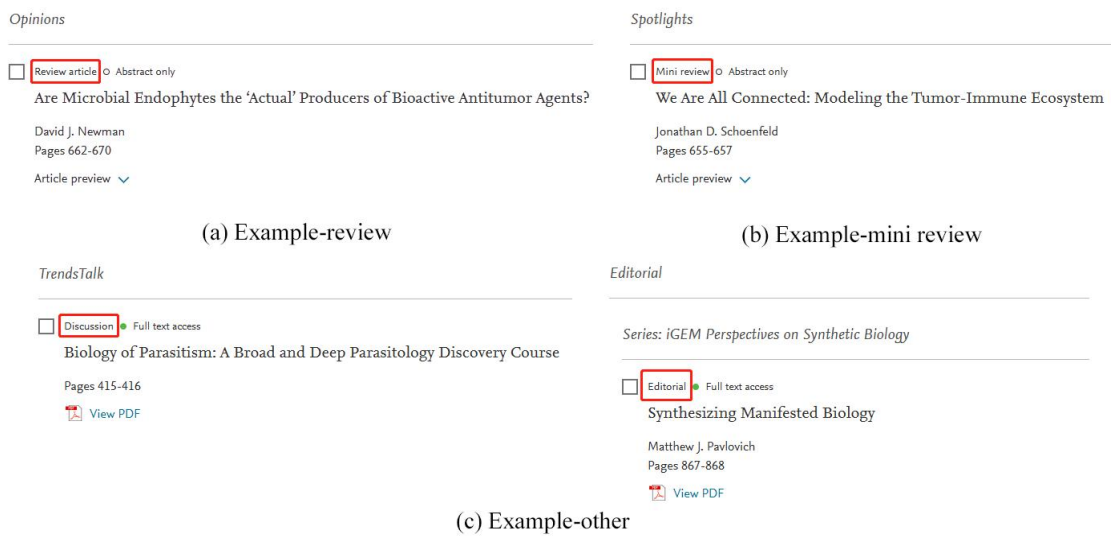

(a) Example-review      (b) Example-mini review

(c) Example-other

**Figure 3.** Examples of document type annotation on the website of Cell Trends In series

Table 3 shows the assignment result for Cell Trends In series. 2121(99.16%) explicit reviews are labeled as "Review" in WoS, and 2137(99.91%) in Scopus. WoS and Scopus have similar accuracy in classifying explicit reviews. Mini reviews are all labeled as "other" in WoS and mainly labeled as "Short Survey"(98.78%) in Scopus. "Short Survey" in Scopus is similar to review, but usually is shorter than traditional review.

**Table 3.** Assignment matrix for Cell Trends series

| Type | Cell Trends | Total | Web of Science | | | | Scopus | | | | |
|---|---|---|---|---|---|---|---|---|---|---|---|
| | | | review | article | other | not indexed | review | article | short survey | other | not indexed |
| Review | All | 2876 | 2121 | 16 | 739 | 0 | 2139 | 2 | 728 | 7 | 0 |
| | Explicit | 2139 | 2121 | 16 | 2 | 0 | 2137 | 2 | 0 | 0 | 0 |
| | Mini Review | 737 | 0 | 0 | 737 | 0 | 2 | 0 | 728 | 7 | 0 |
| Other paper | | 331 | 0 | 0 | 310 | 21 | 0 | 1 | 0 | 330 | 0 |
| Total | | 3207 | 2121 | 16 | 1049 | 21 | 2139 | 3 | 728 | 337 | 0 |

### 3.1.3 Wolters Kluwer Current Opinion journal series

Wolters Kluwer Current Opinion series are pure review journals. Lots of papers are not clearly annotated on the official website. For some editorial papers, the editorial label can only be found in the details pages(Fig.4b). So we check the document type annotation as follows:(1) check the details page of papers and classify papers with "Editorial" on the page as "other paper"; (2)all the rest papers are marked as "explicit review".

**(a)Example-"editorial" included in the title**

**(b) Example-"editorial" not included in the title**

**Figure 4. Document** Type annotation on the website of Current Opinion series

Table 4 shows the result for Wolters Kluwer Current Opinion journals. We can see that only a very small fraction of explicit reviews are misassigned in WoS(0.8%) and Scopus(1.5%). About half of other papers are not indexed in WoS and Scopus.

**Table 4.** Assignment matrix for Wolters Kluwer Current Opinion series

| Type | | Total | Web of Science | | | | Scopus | | | |
|---|---|---|---|---|---|---|---|---|---|---|
| | | | review | article | others | not indexed | review | article | others | not indexed |
| Review | Explicit | 3744 | 3714 | 25 | 0 | 5 | 3696 | 56 | 0 | 0 |
| | Other paper | 589 | 6 | 2 | 302 | 279 | 41 | 12 | 268 | 260 |
| | Total | 4333 | 3720 | 27 | 302 | 284 | 3737 | 68 | 268 | 260 |

### 3.1.4 Review of Modern Physics

*Reviews of Modern Physics* (RMP) is the world's premier physics review journal. But for the papers under investigation, none of explicit reviews are labeled as "Review" in WoS and Scopus as shown in Table 5. One paper published as "Colloquium summary" is labeled as "Review" in Scopus.

**Table 5.** Assignment matrix for *Reviews of Modern Physics*

| Type | | Total | Web of Science | | | | Scopus | | | |
|---|---|---|---|---|---|---|---|---|---|---|
| | | | review | article | others | not indexed | review | article | others | not indexed |
| Review Explicit | | 58 | 0 | 58 | 0 | 0 | 0 | 58 | 0 | 0 |
| Other paper | | 28 | 0 | 23 | 3 | 2 | 1 | 22 | 4 | 1 |
| Total | | 86 | 0 | 81 | 3 | 2 | 1 | 80 | 4 | 1 |

### 3.1.5 WIREs journal series

WIREs clearly divides papers into 6 categories in the website description as shown in Table 6. Papers under the section of "Advanced Review(s)" are classified as "explicit review". Papers under the section of "Focus Article", "Primer", "Overview(s)", "Software Focus", and "Perspective" are classified as "implicit review" .

**Table 6.** Official website descriptions of the main types for WIREs series

| Website Type | Website Description | Mapping Type |
|---|---|---|
| Advanced Review | *These articles review key areas of research in a citation-rich format similar to that of leading review journals.* | explicit review |
| Focus Article | *These articles are mini-reviews, and which therefore illustrate aspects of larger ideas covered in Overviews and Advanced Reviews.* | implicit review |
| Primer | *Meant to be understood by a very general audience. These articles should provide orientation to the key theories, knowledge, uncertainties, and controversies in the field.* | implicit review |
| Overview | *Broad and relatively non-technical treatment of important topics at a level. These articles must refer to the key articles/books in the field (not exhaustive but comprehensive).* | implicit review |
| Software Focus | *These articles should review the capabilities of the software and how it has been and can be applied.* | implicit review |
| Perspective | *A forum for thought-leaders, hand-picked. They should cite literature which authenticates their argument(s), but without the need to be exhaustive or comprehensive.* | implicit review |

Table 7 shows the result for WIREs journals. Most of explicit reviews are assigned as "Review" in WoS(99%) and Scopus(87%), while more than 50% of the indexed implicit reviews are mislabeled in WoS and Scopus. The classification accuracy of explicit reviews is much better than that of implicit reviews. This phenomenon occurs probably because implicit section names provide some confounding information which makes judgment more difficult.

In WoS, 238 implicit reviews are labeled as "Article", including 114 "focus article", 84 "overview(s)", 19 "perspective", 16 "primer", and 5 "software focus". Among 150 reviews labeled as "Article" in Scopus, there are 70 "focus article", 43 "overview", 15 "perspective",

14 "advanced review(s)", 4 "primer" and 4 "software focus". In conclusion, "Focus Article" is the most commonly misclassified section, probably because of its confusing section name.

**Table 7.** Assignment matrix for WIREs series

| Type | | Total | Web of Science | | | | Scopus | | | |
|---|---|---|---|---|---|---|---|---|---|---|
| | | | review | article | others | not indexed | review | article | others | not indexed |
| Review | All | 731 | 487 | 241 | 3 | 0 | 466 | 150 | 0 | 115 |
| | Explicit | 386 | 383 | 3 | 0 | 0 | 336 | 14 | 0 | 36 |
| | Implicit | 345 | 104 | 238 | 3 | 0 | 130 | 136 | 0 | 79 |
| Other paper | | 201 | 1 | 12 | 10 | 178 | 6 | 8 | 9 | 178 |
| Total | | 932 | 488 | 253 | 13 | 178 | 472 | 158 | 9 | 293 |

## 3.2 Descriptive results of Review assignment for mixed review journals

### 3.2.1 Elsevier Current Opinion series

As for Elsevier Current Opinion journals, section heading is contained in the content page of corresponding volume and there are 16 section heading types totally. Papers under the section of "Review Article" are marked as "explicit review" and papers of "Research articles" will be divided into corresponding types according to their abstract and full-text. Papers under the other 14 sections (e.g., "Erratum", "Correspondence") are classified as "other paper".

The assignment matrix is shown in Table 8. Towarding the assignment of reviews, Scopus performs better than WoS generally. In WoS, 1125 review papers (26.5%) are labelded as "Article", and this mislabeling mainly happens for explicit reviews (1121/1125). In addition there are 22 explicit reviews assigned as "other paper" and 4 of the 13 implicit reviews are assigned as "Article" in WoS. The assignment of several journals is extremely problematic, e.g, 197/197 of reviews in *Current Opinion in Virology*, 249/249 in *Current Opinion in Structural Biology* (249/249), and 206/208 in *Current Opinion in Pharmacology* are labeled as "Article".

While the mislabeling proportion of explicit reviews in Scopus is low (0.024%) compared to WoS, all the implicit reviews are misassigned in Scopus. The high consistency between Current Opinions and Scopus may be due to they both belonging to Elsevier.

**Table 8.** Assignment matrix for Elsevier Current Opinion series

| Type | | Total | Web of Science | | | | Scopus | | | |
|---|---|---|---|---|---|---|---|---|---|---|
| | | | review | article | others | not indexed | review | article | others | not indexed |
| Review | All | 4238 | 3089 | 1125 | 22 | 2 | 4224 | 13 | 0 | 1 |

| | | | | | | | | | |
|---|---|---|---|---|---|---|---|---|---|
| | Explicit | 4225 | 3080 | 1121 | 22 | 2 | 4224 | 0 | 0 | 1 |
| | Implicit | 13 | 9 | 4 | 0 | 0 | 0 | 13 | 0 | 0 |
| Article | Implicit | 3 | 3 | 0 | 0 | 0 | 0 | 3 | 0 | 0 |
| Other paper | | 742 | 3 | 0 | 354 | 385 | 0 | 2 | 365 | 375 |
| Total | | 4983 | 3095 | 1125 | 376 | 387 | 4224 | 18 | 365 | 376 |

### 3.2.2 Nature Reviews series

As for Nature Reviews, there are 12 subtypes. Papers under the section of "Review" are marked as "explicit review" and papers under the "Research" section are classified according to the full-text. Papers under sections like "Research Highlights", "Editorial", "News & Views", are marked as "other paper".

Table 9 shows the assignment matrix for Nature Reviews sereis. We can see that 201(11.24%) explicit reviews are marked as "Article" in WoS and 115(6.43%) in Scopus. There are 90 explicit reviews in WoS and 57 in Scopus are not indexed. Another interesting observation can be found for the document type assignment of "other paper" in Scopus. We can see that about 1,200 papers as asigned as review, article, and short survey. A detailed distribution can be found in Fig.5. Among the 301 papers labeled as short survey, most of them are from the website section "news & views". The "article" papers are mainly from "research highlight".

**Table 9.** Assignment matrix for Nature Reviews series

| Type | | Total | Web of Science | | | | Scopus | | | | |
|---|---|---|---|---|---|---|---|---|---|---|---|
| | | | review | article | others | not indexed | review | article | short survey | others | not indexed |
| Review | All | 1799 | 1501 | 207 | 1 | 90 | 1589 | 123 | 0 | 30 | 57 |
| | Explicit | 1788 | 1496 | 201 | 1 | 90 | 1586 | 115 | 0 | 30 | 57 |
| | Implicit | 11 | 5 | 6 | 0 | 0 | 3 | 8 | 0 | 0 | 0 |
| Other paper | | 4178 | 3 | 11 | 3346 | 818 | 127 | 784 | 301 | 2770 | 196 |
| Total | | 5977 | 1504 | 218 | 3347 | 908 | 1716 | 907 | 301 | 2800 | 253 |

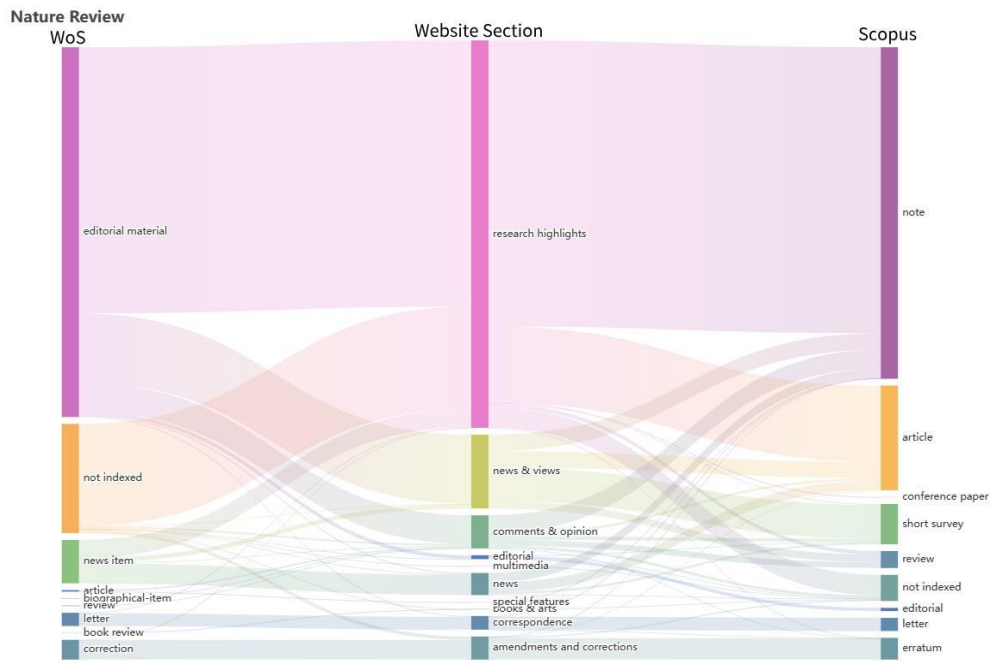

**Figure 5. Distribution of document types for other papers on website, WoS and Scopus.**

In the Nature Reviews, there is one special journal -- *Nature Reviews Disease Primers*. Each explicit review of *NRDP* contains two versions -- "Primers" and "PrimerViews". "Primer" is an introductory review article, and "PrimerViews" is an infographic that accompanies each Primer article showing the central message to patients in the form of visual summaries (Figure 6). All the "Primers"(90) from *NRDP* are labeled as "Article" and all the "PrimerViews"(90) are not included in WoS. While some "PrimerViews"(56.67%) from *NRDP* are included in Scopus.

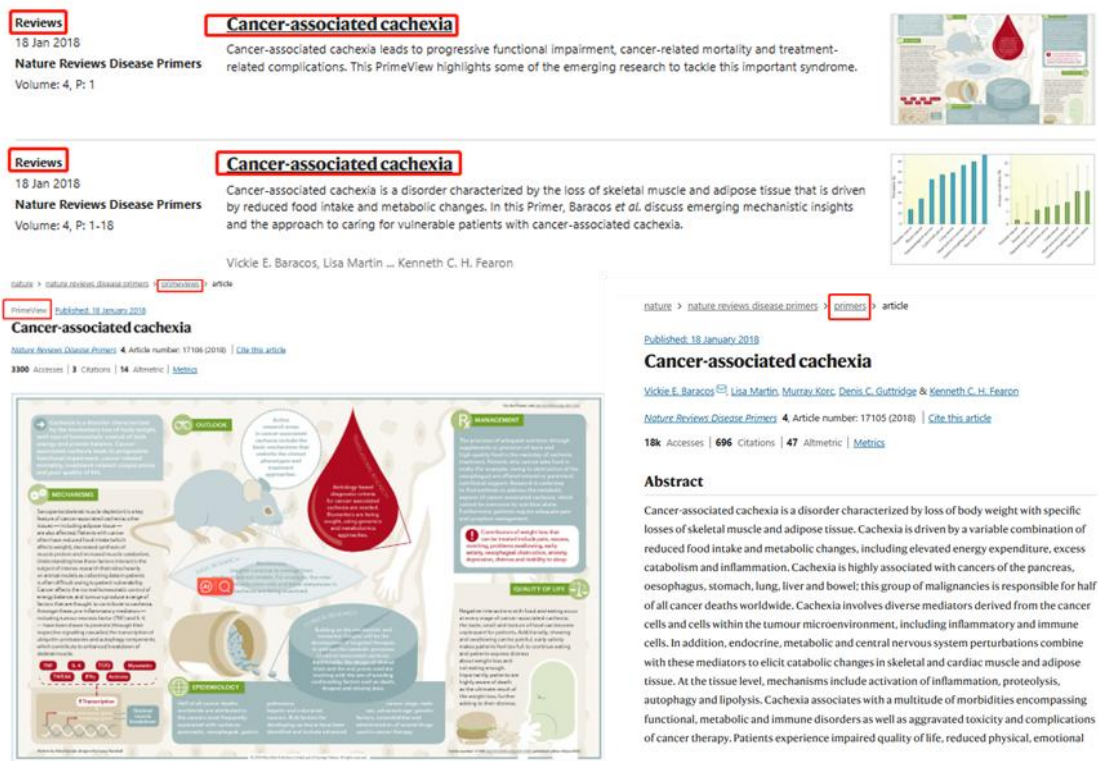

**Figure 6.** Example of PrimerViews and Primer in *Nature Reviews Disease Primers.*

### 3.2.3 Taylor & Francis Expert Opinion Series

Each article has two levels of section heading in the corresponding volume's content list page as shown in Fig.7. The first-level headings have 34 names on the official websites. Despite these three headings --"Reviews", "Clinical focus: rare blood disorders - Review" and "Clinical features - Review", the other 31 headings do not have clear words indicating its document type. In addition, the secondary headings are very inaccurate (Fig.7a). Therefore, we map these headings based on the website description. For example, official website describes "Special Report" as "*short review-style articles that summarize a particular niche area, be it a specific technique or therapeutic method*", so papers under this section are put into "implicit review".

(a) Example-implicit review

(b) Example-explicit review

**Figure 7. The annotation in website of Taylor & Francis Expert Opinion Series.**

The assignment matrix for Taylor & Francis Expert Opinion journals is shown in Table 10. 86.59% explicit reviews are annotated as "Review" and 41.33% implicit reviews are marked as "Review" in WoS. The above two proportions are 98.79% and 9.33% in Scopus. The accuracy of explicit reviews marking is much better than implicit reviews marking and this phenomenon also appears in the task of research article assignment. Scopus performs slightly better in the task of assigning explicit review, while WoS is relatively better in the accuracy of assigning implicit review in this dataset.

**Table 10.** Assignment matrix for Taylor & Francis Expert Opinion

| Type | Taylor & Francis Expert Opinion | Total | Web of Science | | | | Scopus | | | |
|---|---|---|---|---|---|---|---|---|---|---|
| | | | review | article | others | not indexed | review | article | others | not indexed |
| Review | All | 2106 | 1620 | 453 | 33 | 0 | 1677 | 427 | 2 | 0 |
| | Explicit | 1656 | 1434 | 221 | 1 | 0 | 1635 | 19 | 2 | 0 |
| | Implicit | 450 | 186 | 232 | 32 | 0 | 42 | 408 | 0 | 0 |
| Article | Total | 155 | 48 | 105 | 2 | 0 | 7 | 147 | 1 | 12 |

| | Total | review | article | others | not indexed | review | article | others | not indexed |
|---|---|---|---|---|---|---|---|---|---|
| Explicit | 12 | 6 | 5 | 1 | 0 | 6 | 5 | 1 | 12 |
| Implicit | 143 | 42 | 100 | 1 | 0 | 1 | 142 | 0 | 0 |
| Other paper | 258 | 2 | 0 | 256 | 0 | 1 | 2 | 255 | 0 |
| Total | 2519 | 1670 | 558 | 291 | 0 | 1685 | 576 | 258 | 0 |

### 3.2.4 Taylor & Francis Expert Review Series

Taylor & Francis Expert Review Series have a quite similar website schema as Taylor & Francis Expert Opinion Series. Table 11 shows the result for Taylor & Francis Expert Review journals. A similar pattern can be found in Table 11 as Table 10: for explicit reviews, Scopus performs better than WoS; for implicit review, WoS performs slightly better.

**Table 11.** Assignment matrix for Taylor & Francis Expert Review

| Type | | Total | Web of Science | | | | Scopus | | | |
|---|---|---|---|---|---|---|---|---|---|---|
| | | | review | article | others | not indexed | review | article | others | not indexed |
| Review | All | 2151 | 1791 | 358 | 2 | 0 | 1827 | 319 | 5 | 0 |
| | Explicit | 1815 | 1698 | 117 | 0 | 0 | 1814 | 1 | 0 | 0 |
| | Implicit | 336 | 93 | 241 | 2 | 0 | 13 | 318 | 5 | 0 |
| Article | Explicit | 231 | 27 | 203 | 1 | 0 | 3 | 228 | 0 | 0 |
| Other paper | | 355 | 2 | 0 | 353 | 0 | 1 | 1 | 353 | 0 |
| Total | | 2737 | 1820 | 561 | 356 | 0 | 1831 | 548 | 358 | 0 |

241 implicit reviews labeled as "Article" are mainly from "Drug Profile", "Perspective" and "Special Report" in WoS(51.45%, 20.75%, 14.11%) and Scopus(43.71%, 22.96%, 21.07%). "Drug Profile" and "Special Report" review some experimental methods or experimental data, and authors will present criticism or address controversy in "Perspective". It could interfere with judgment of databases on the content.

### 3.2.5 Taylor & Francis Critical Reviews

As for Taylor & Francis Critical Reviews, papers under the section of "Review Article", "Review" and "Critical Review" are regarded as "explicit review" and papers under "Article(s)", "Original Article(s)" and "Short Article" are marked as "implicit review".

As shown in Table 12, WoS and Scopus both perform well in annotating explicit reviews. Almost all implicit reviews were marked as "Review" in WoS. For the implicit reviews in Taylor & Francis Critical Reviews, WoS performs better than Scopus.

**Table 12.** Assignment matrix for Taylor & Francis Critical Reviews

| Type | | Total | Web of Science | | | | Scopus | | | |
|---|---|---|---|---|---|---|---|---|---|---|
| | | | review | article | others | not indexed | review | article | others | not indexed |
| Review | All | 1148 | 679 | 0 | 1 | 0 | 701 | 445 | 2 | 0 |

| | | | | | | | | | |
|---|---|---|---|---|---|---|---|---|---|
| Explicit | 627 | 627 | 0 | 0 | 0 | 619 | 7 | 1 | 0 |
| Implicit | 521 | 520 | 0 | 1 | 0 | 82 | 438 | 1 | 0 |
| Other paper | 32 | 2 | 0 | 30 | 0 | 4 | 2 | 24 | 2 |
| Total | 1180 | 1149 | 0 | 31 | 0 | 705 | 447 | 26 | 2 |

## 3.3 Overview of assignment performance for these review journal series

In the above sections, we illustrated the comparison of document type assignment across website, WoS and Scopus for each review journal series. Here we summarize the assignment performance of WoS and Scopus as show in Fig.8.

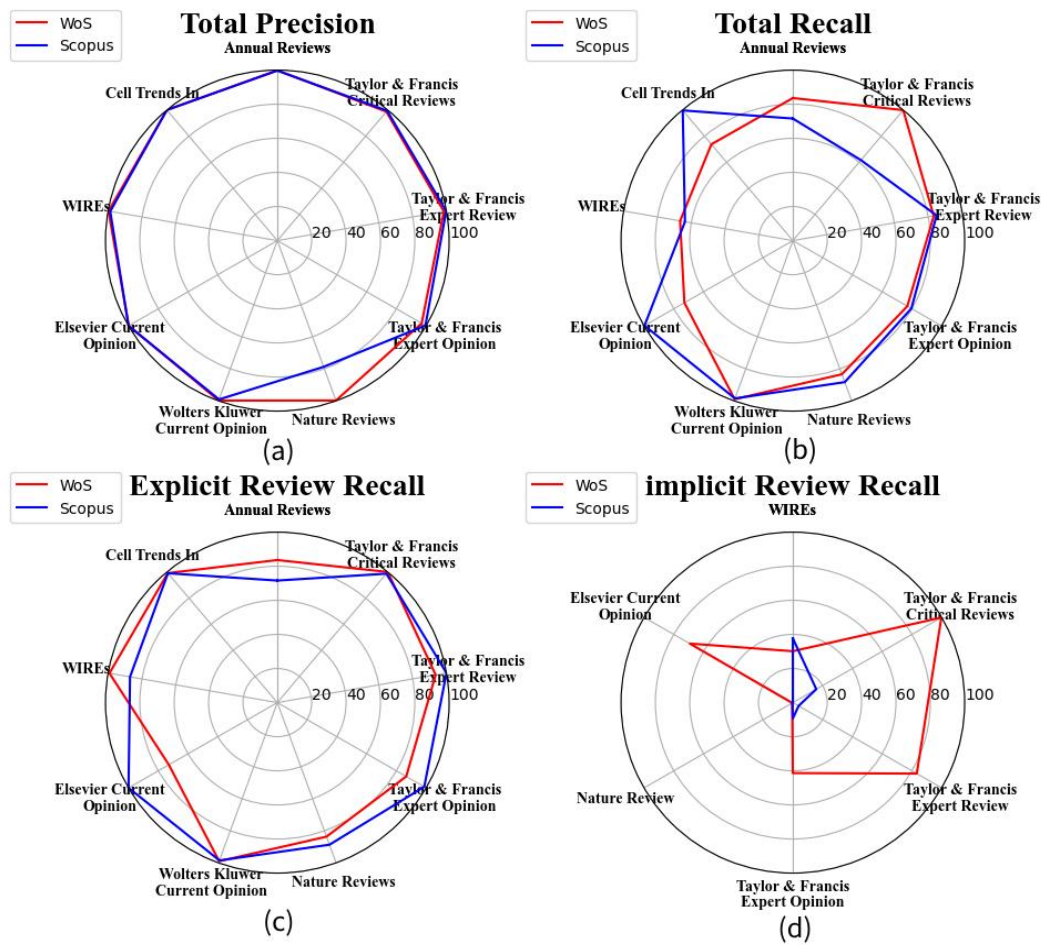

**Figure 8. Assignment precision and recall of review articles.** (a)-(d) respectively show the total precision, total recall, explicit review recall and implicit review recall. (d) just represent the results of 6 mixed review journal series.

Figure 8(a) and Figure 8(b) respectively represents the results of total precision (including explicit review and implicit review) and total recall for each journal series. In general, WoS and Scopus have high performance toward total precision(exceeding 97%) in most journal series. However, the total precision of RMP is 0% in WoS and Scopus, and the precision of

Nature Reviews series is 78.78% in Scopus. WoS and Scopus show some difference in term of recall. For example, Scopus performs better for Cell Trends In series and Current Opinion series, while WoS performs better for Annual Reviews and Taylor & Francis Critical Reviews. When the recall of explicit reviews and implicit reviews are displayed seperately, we can see a huge difference. Compared with the recall of explicit reviews shown in Fig.8(c) , the recall of implicit reviews shown in Fig.8(d) is much smaller, and the performance of WoS is better than Scopus. This observation implies that the two databases should pay more attention to these implicit reviews in document type assignment, and publishers can better assigning the labels on their websites.

## 4. Conclusion and Discussion

In the present study, 160 review journals of ten brand series are selected to investigate the document-type assigning accuracy of review articles in WoS and Scopus. The document type annotated on the official website is treated as the golden-standard. We further classified these reviews as explicit and implicit based on whether the section heading or the online annotation directly indicating it is review or not. Overall, WoS and Scopus performed similarly, with an average precision of about 99% and recall of about 80%. However, there were some differences between WoS and Scopus across different journal series. In some series (e.g. Cell Trends In), Scopus performed better, while in other series (e.g., the Critical Review series), WoS performed better. After differentiating between explicit reviews and implicit reviews, we can see that the assigning accuracy of WoS and Scopus for implicit reviews dropped significantly, especially for Scopus. These two databases need to devote more effort to correctly labeling the document types of implicit reviews, and the publishers may annotate document type on the website more clearly. In addition, when we looked deeper into the labeling of document types within journal series, we found huge differences in labeling accuracy even among journals belonging to the same series, with some journals being completely mislabeled. To address this issue, we recommend WoS and Scopus identify these journals and unifying the document type labeling across them.

This study has some limitations that need to be considered when interpreting its results. Firstly, the document types we used as the gold standard were based on the journal websites' labeling, and we did not manually validate each paper based on full text, so there may be some accidental mislabeling. Secondly, in this paper, we only studied the labeling performance for review articles published in review journals. Whether this conclusion can be extended to review articles published in non-review journals is not very clear. However, compared to previous studies, the recall is fairly consistent. Thirdly, the papers analyzed in this study were published during 2017-2018, and may not fully reflect the most current situation. In addition, there is currently no universally agreed-upon definition for review articles. Papers like mini reviews, perspective papers, commentaries, greatly increase the difficulty of document type labeling. Compared to WoS, Scopus has an additional "Short Survey" document type, which may be one option to solve this problem.

Here are some suggestions for future work: 1) Analyze the document type assignment for reviews across different research fields. There are some differences in how reviews are written and used across disciplines, for example meta-analyses and systematic reviews are very common in medicine. 2) Examine labeling of review articles published in regular journals. This paper only analyzed review articles published in review journals, which account for just a portion of all review articles, and do not fully reflect overall database coverage. 3) Use state-of-the-art AI methods to assist with labeling reviews in order to improve assigning accuracy.

# Acknowledgements

We'd like to thank Mike Jones(the sales manager of Annual Reviews), Xiaolin Li(Wiley Open Research and Journal Development Manager), and Jason Hu(the Senior Vice President of Taylor & Francis) for the prompt reply about the specific type of documents published in the review journals. We would like thank Dr.Weiping Yue for the description of criteria of review used in WoS.

# Author contributions

Manman Zhu: Writing - Original Draft; Data Curation; Investigation
Xinyue Lu: Data Curation; Investigation; Visualization; Writing - Original Draft
Fuyou Chen: Data Curation
Liying Yang: Writing - Review & Editing
Zhesi Shen: Conceptualization; Data Curation; Investigation; Writing - Original Draft; Writing - Review & Editing

# Data availability

The data analyzed in this study and the html files for the figures shown in appendix can be accessed via https://www.scidb.cn/en/s/neuYVb.

# Appendix

The correspondence of document types on website, WoS and Scopus for the journal series analyzed in this study is shown the following figures respectively. In each figure, the left column shows the document types in WoS, the middle column shows the document type on publishers' websites, and the right column shows the document type in Scopus. In the middle column, for each type we show the aggregated type in the parenthesis.

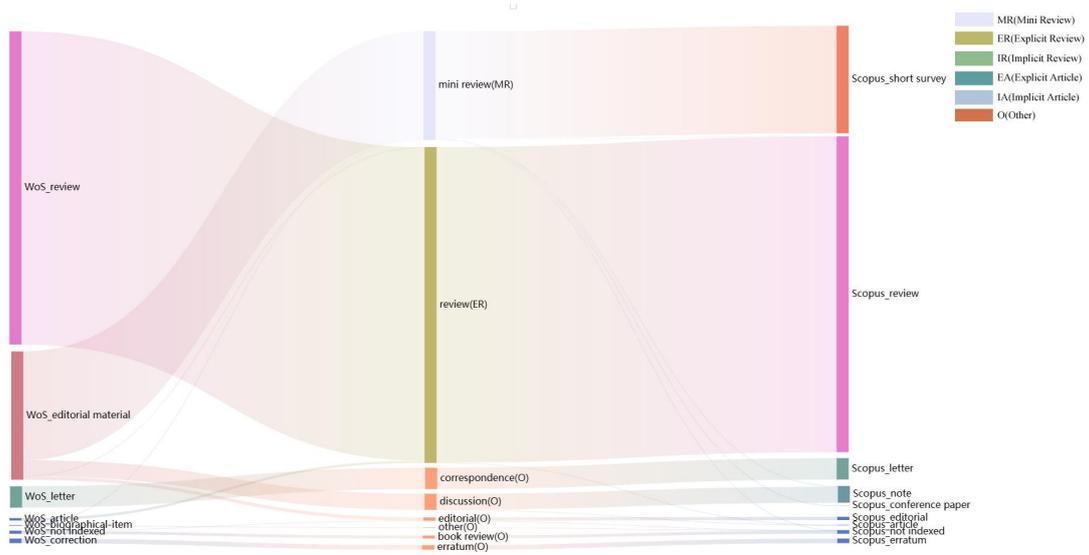

**Figure A1. Correspondence of document types on website, WoS and Scopus for Cell Trends In journal series.**

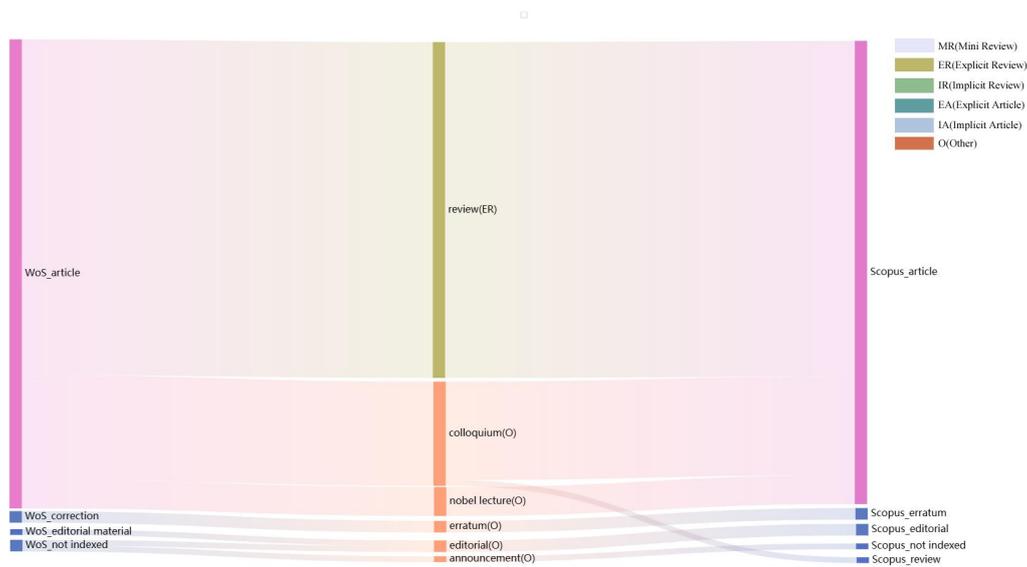

**Figure A2. Correspondence of document types on website, WoS and Scopus for Review of Modern Physics.**

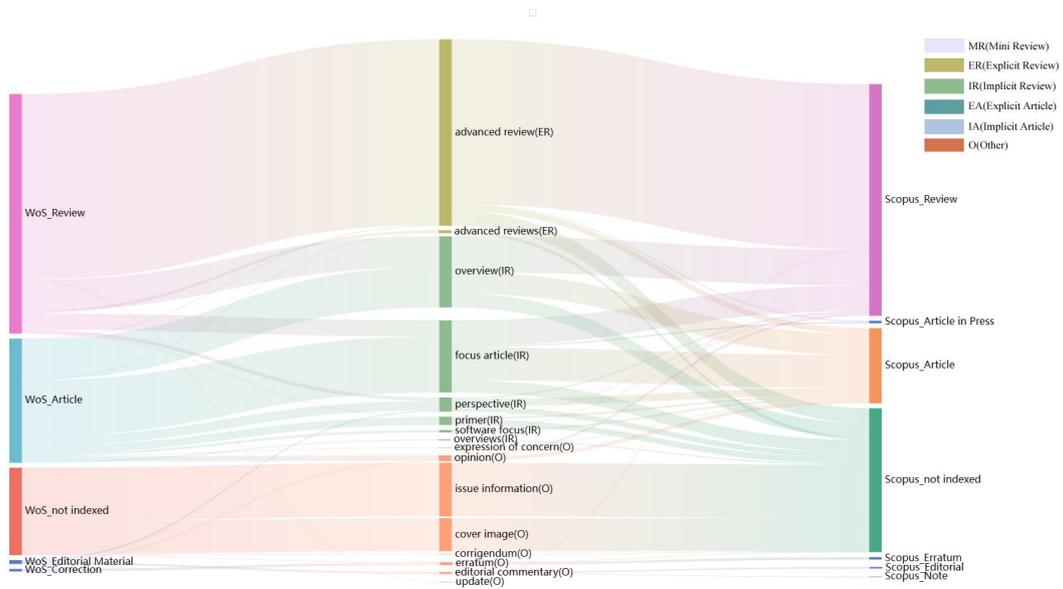

**Figure A3. Correspondence of document types on website, WoS and Scopus for WIREs journal series.**

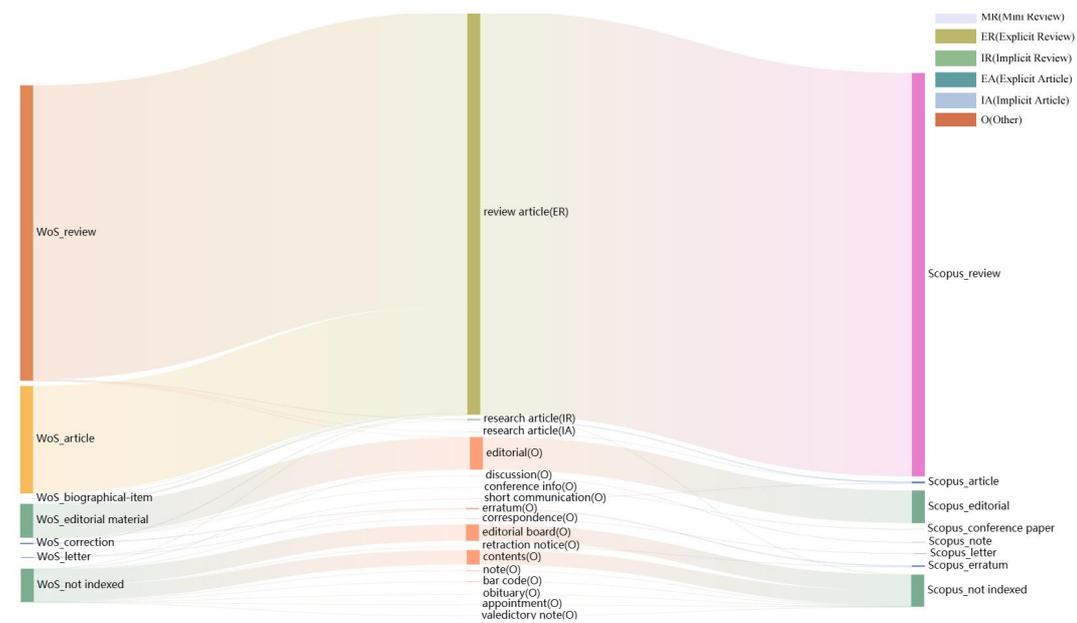

**Figure A4. Correspondence of document types on website, WoS and Scopus for Elsevier Current Opinion series .**

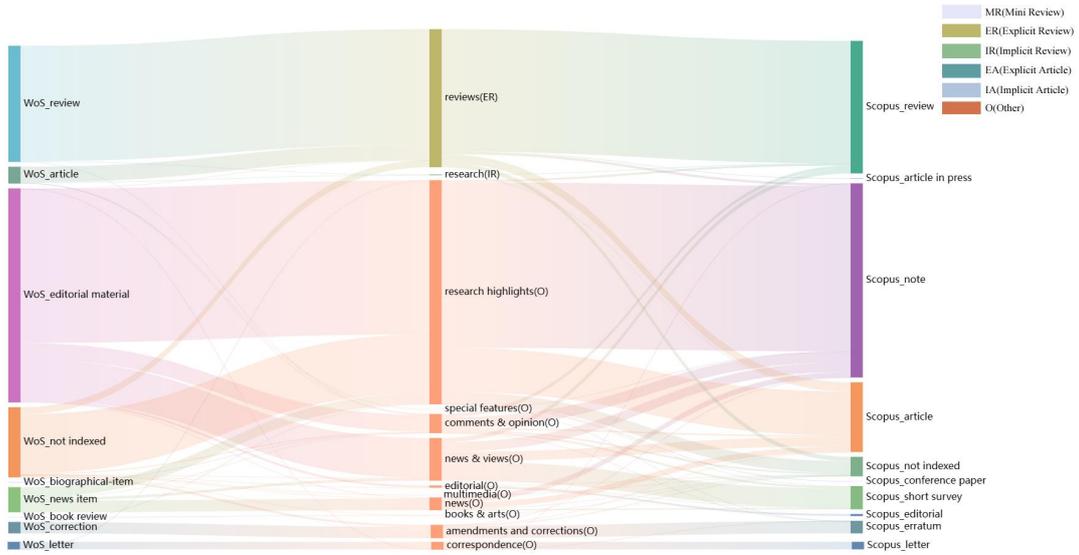

**Figure A5. Correspondence of document types on website, WoS and Scopus for Nature Reviews series.**

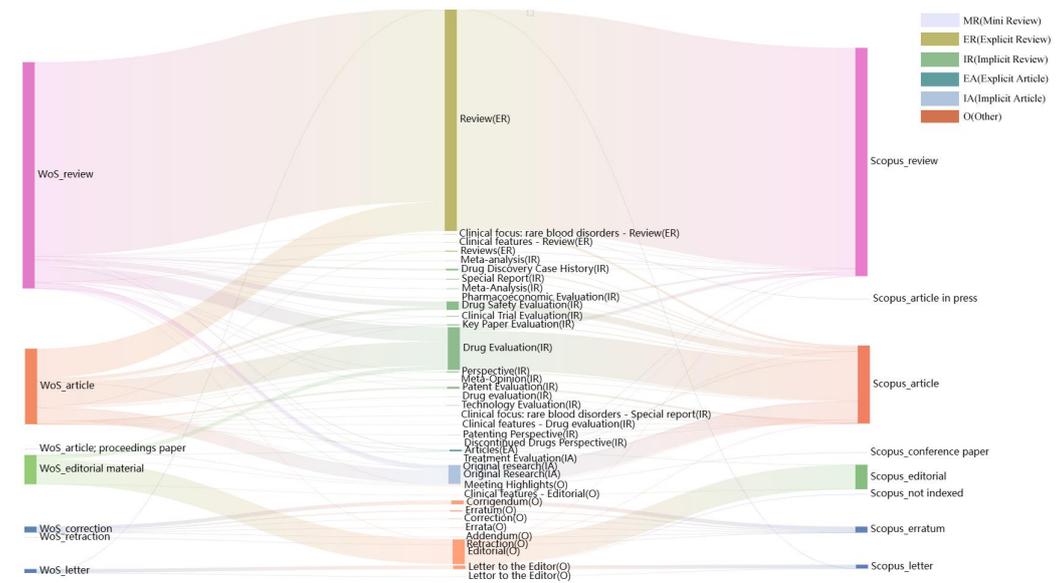

**Figure A6. Correspondence of document types on website, WoS and Scopus for Taylor & Francis Expert Opinion Series .**

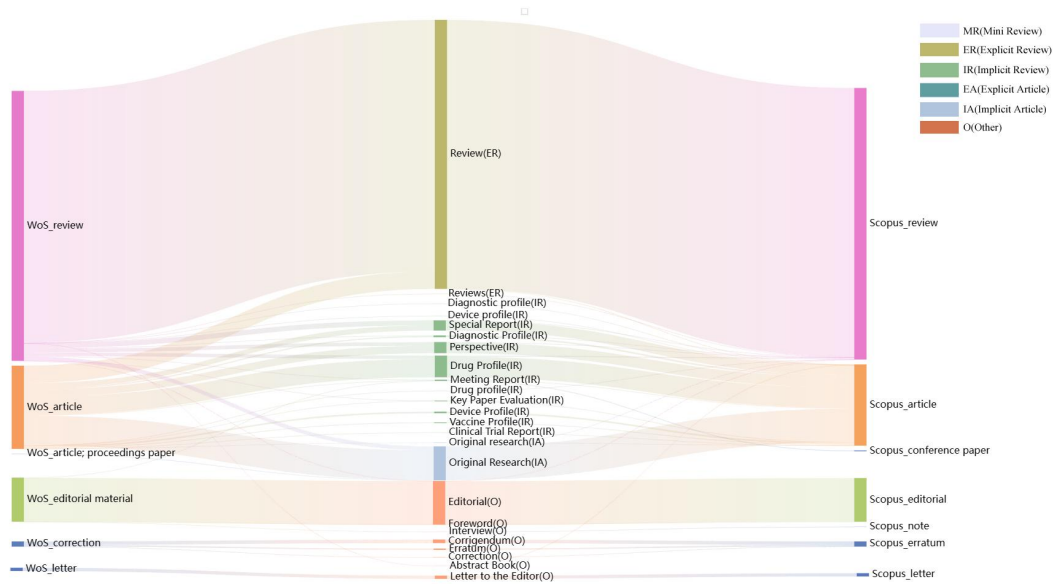

**Figure A7. Correspondence of document types on website, WoS and Scopus for Taylor & Francis Expert Review Series .**

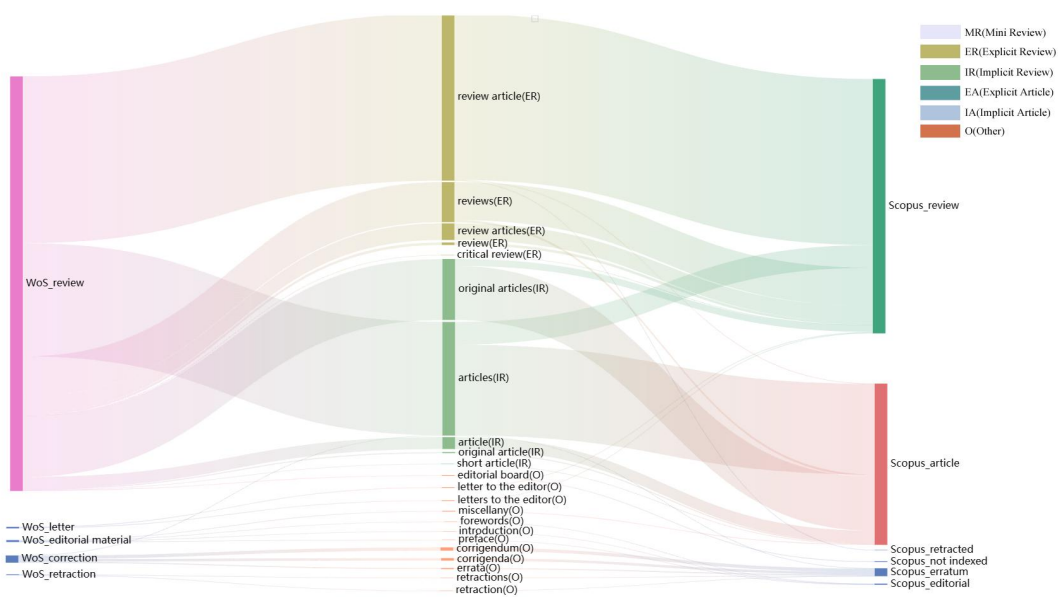

**Figure A8. Correspondence of document types on website, WoS and Scopus for Taylor & Francis Critical Reviews.**